\documentclass[a4paper,11pt]{article}
\pdfoutput=1 

\usepackage{jheppub} 

\usepackage[T1]{fontenc} 
\usepackage{slashed}
\usepackage[normalem]{ulem}
\usepackage[makeroom]{cancel}

\title{\boldmath BPS Equations of Monopole and Dyon in $SU(2)$ Yang-Mills-Higgs Model, Nakamula-Shiraishi Models, and Their Generalized Versions from The BPS Lagrangian Method}

\author[1]{Ardian Nata Atmaja,\note{Corresponding author.}}
\author[1,2]{Ilham Prasetyo}

\affiliation[1]{Research Center for Physics, Indonesian Institute of Sciences (LIPI),\\Kompleks PUSPIPTEK Serpong, Tangerang 15310, Indonesia}
\affiliation[2]{Departemen Fisika, FMIPA, Universitas Indonesia,\\Depok 16424, Indonesia}

\emailAdd{ardi002@lipi.go.id}
\emailAdd{ilham.prasetyo@sci.ui.ac.id}

\abstract{We apply the BPS Lagrangian method~\cite{Atmaja:2015umo} to derive BPS equations of monopole and dyon in the $SU(2)$ Yang-Mills-Higgs model, Nakamula-Shiraishi models, and their Generalized versions. We argue that by identifying the effective fields of scalar field, $f$, and of time-component gauge field, $j$, explicitly by $j=\beta f$ with $\beta$ is a real constant, the usual BPS equations for dyon can be obtained naturally. We validate this identification by showing that both Euler-Lagrange equations for $f$ and $j$ are identical in the BPS limit. The value of $\beta$ is bounded to $|\beta|<1$ due to reality condition on the resulting BPS equations. In the Born-Infeld type of actions, namely Nakamula-Shiraishi models and their Generalized versions, we find a new feature that adding the energy density by a constant $4b^2$, with $b$ is the Born-Infeld parameter, will turn monopole(dyon) to anti-monopole(anti-dyon) and vice versa. In all Generalized versions there are additional constraint equations that relate the scalar-dependent couplings of scalar and of gauge kinectic terms; or $G$ and $w$ respectively. For monopole the constraint equation is $G=w^{-1}$, while for dyon is $w(G-\beta^2 w)=1-\beta^2$ which further gives lower bound to $G$ as such $G\geq|2\beta\sqrt{1-\beta^2}|$. We also write down the complete square-forms of all effective Lagrangians.}

\begin{document} 
\maketitle
\flushbottom

\section{Introduction}

Monopole has been known to exist in non-abelian gauge theory. One of the main developments was given by 't Hooft in~\cite{tHooft:1974kcl}, and in parallel with a work by Polyakov in~\cite{Polyakov:1974ek}, in which he showed that monopole could arise as soliton in a Yang-Mills-Higgs theory, without introducing Dirac's string~\cite{Dirac:1931kp}, by spontaneously breaking the symmetry of $SO(3)$ gauge group into $U(1)$ gauge group. Later on, Julia and Zee showed that a more general configuration of soliton called dyon may exist as well within the same model~\cite{Julia:1975ff}. Futhermore, the exact solutions were given by Prasad and Sommerfiled in~\cite{Prasad:1975kr} by taking some limit where $V\to 0$. These solutions were proved by Bogomolnyi in~\cite{Bogomolny:1975de} to be solutions of the first-order differential equations which turn out to be closely related with the study of supersymmetric system~\cite{Witten:1978mh}\footnote{In this article, we shall call the limit $V\to0$ as BPS limit and the first-order differential equations as BPS equations.}.

At high energy the Yang-Mills theory may receive contributions from higher derivative terms. This can be realized in string theory in which the effective action of open string theory may be described by the Born-Infeld type of actions~\cite{Abouelsaood:1986gd}. However, there are several ways in writing the Born-Infeld action for non-abelian gauge theory because of the ordering of matrix-valued field strength~\cite{Abouelsaood:1986gd,Tseytlin:1997csa,Hashimoto:1997gm,Gross:1997mx,Brecher:1998tv,Gonorazky:1998ay}. Further complications appear when we add Higgs field into the action. One of examples has been given by Nakamula and Shiraishi in which the action exhibits the usual BPS monopole and dyon~\cite{Nakamula:2013rfa}. Unfortunately, the resulting BPS equations obviously do not capture essensial feature of the Born-Infeld action namely there is no dependency over the Born-Infeld parameter. Other example such as in~\cite{Grandi:1999rv}, the monopole's profile depends on the Born-Infeld parameter, but the BPS equations are not known so far.

In this article, we would like to derive the well-known BPS equations of monopole and dyon in the $SU(2)$ Yang-Mills-Higgs model and their Born-Infeld type extensions, which we shall call them Nakamula-Shiraishi models, using a procedure called BPS Lagrangian method developed in~\cite{Atmaja:2015umo}. We then extend those models to their Generalized versions by adding scalar-dependent couplings to each of the kinetic terms and derive the BPS equations for monopole and dyon. In section 2, we will first discuss in detail about the BPS Lagrangian method. The next section 3, we describe how to get the BPS equations for monopole(dyon) from energy density of the $SU(2)$ Yang-Mills-Higgs model using Bogomolny's trick. We write explicitly its effective action and effective actions of the Nakamula-Shiraishi models by taking the 't Hooft-Polyakov(Julia-Zee) ansatz for monopole(dyon). In section 4, we use the BPS Lagrangian approach to reproduce the BPS equations for monopole and dyon in the $SU(2)$ Yang-Mills-Higgs model and Nakamula-Shiraishi models. Later, in section 5, we generalize the $SU(2)$ Yang-Mills-Higgs model by adding scalar-dependent couplings to scalar and gauge kinetic terms and derive the corresponding BPS equations. We also generalize Nakamula-Shiraishi models in section 6 and derive their corresponding BPS equations. We end with discussion in section 7.

\section{BPS Langrangian Method}
In deriving BPS equations of a model, we normally use so called Bogomolnyi's trick by writing the energy density into a complete square form~\cite{Bogomolny:1975de}. However, there are several rigorous methods have been developed in doing so. The first one is based on the Bogomolnyi's trick by assuming the existance of a homotopy invariant term in the energy density that does not contribute to Euler-Lagrange equations~\cite{Adam:2013hza}. The second method called first-order formalism works by solving a first integral of the model, together with stressless condition,~\cite{Bazeia:2005tj,Bazeia:2006mh,Bazeia:2007df}. The third method called On-Shell method works by adding and solving auxiliary fields into the Euler-Lagrange equations and assuming the existance of BPS equations within the Euler-Lagrange equations~\cite{Atmaja:2014fha,Atmaja:2015umo}. The forth method called First-Order Euler-Lagrange(FOEL) formalism, which is generalization of Bogomolnyi decomposition using a concept of strong necessary condition developed in~\cite{Sokalski:2001wk}, works by adding and solving a total derivative term into the Lagrangian~\cite{Adam:2016ipc}\footnote{In our opinion the procedure looks similar to the On-Shell method by means that adding total derivative terms into the Lagrangian is equivalent to introducing auxiliary fields in the Euler-Lagrange equations. However, we admit that the procedure is written in a more covariant way.}. The last method, which we shall call BPS Lagrangian method, works by identifying the (effective)Lagrangian with a BPS Lagrangian such that its solutions of the first-derivative fields give out the desired BPS equations~\cite{Atmaja:2015umo}. This method was developed based on the On-Shell method by one of the author of this article and it is much easier to execute compared to the On-Shell method. We chose to use the BPS Lagrangian method to find BPS equations of all models considered in this article. The method is explained in the following paragraphs. 

In general the total static energy of N-fields system, $\vec{\phi}=(\phi_1,\ldots,\phi_N)$, with Lagrangian density $\mathcal{L}$ is defined by $E_\text{static}=-\int d^d x~ \mathcal{L}$. The Bogomolnyi's trick explains that the static energy can be rewritten as 
\begin{equation}
E_\text{static}=\left(\int d^dx \sum_{i=1}^N \Phi_i(\vec{\phi},\partial\vec{\phi})\right)+E_\text{BPS}, \label{Bogomolnyi trick}
\end{equation}
with $\{\Phi_i\}$ is a set of positive-semidefinite functions and $E_\text{BPS}$ is the boundary contributions defined by $E_\text{BPS}=-\int d^dx~ \mathcal{L}_\text{BPS}$. Neglecting the contribution from boundary terms in $\mathcal{L}_\text{BPS}$, as they do not affect the Euler-Lagrange equations, configurations that minize the static energy are also solutions of the Euler-Lagrange equations and they are given by $\{\Phi_i =0\}$ known as BPS equations. Rewriting the static energy to be in the form of equation (\ref{Bogomolnyi trick}) is not always an easy task. However it was argued in~\cite{Atmaja:2015umo} that one does not need to know the explicit form of equation (\ref{Bogomolnyi trick}) in order to obtain the BPS equations. By realizing that in the BPS limit, where the BPS equations are assumed to be exist, remaining terms in the total static energy are in the form of boundary terms, $E_\text{static}=E_\text{BPS}$. Therefore we may conclude that BPS equations are solutions of $\mathcal{L}-\mathcal{L}_{BPS}=\sum_{i=1}^N \Phi_i(\vec{\phi},\partial\vec{\phi})=0$.

Now let us see in detail what is inside $\mathcal{L}_\text{BPS}$. Suppose that in spherical coordinates the system effectively depends on only radial coordinate $r$. As shown by the On-Shell method on models of vortices~\cite{Atmaja:2014fha}, the total static energy in the BPS limit can be defined as
\begin{equation}
E_\text{BPS}=Q(r\rightarrow\infty)-Q(r\rightarrow 0)=\int^{r\rightarrow\infty}_{r\rightarrow 0} dQ,
\end{equation}
where $Q$ is called BPS energy function. The BPS energy function $Q$ does not depend on the coordinate $r$ explicitly however in general it can also depend on $r$ explicitly in accordance with the choosen ansatz. In most of the cases if we choose the ansatz that does not depend explicitly on coordinate $r$ then we would have $Q\neq Q(r)$. Hence, with a suitable ansatz, we could write $Q=Q(\tilde{\phi}_1,\ldots,\tilde{\phi}_N)$ in which $\tilde{\phi}_i$ is the effective field of $\phi_i$ as a function of coordinate $r$ only. Assume that $Q$ can be treated with separation of variables
\begin{equation}
Q\equiv \prod_{i=1}^N Q_i(\tilde{\phi}_i), \label{eq:guessQ}
\end{equation}
this give us a pretty simple expression of $E_\text{BPS}$, i.e.
\begin{equation}
E_\text{BPS}=\int \sum_{i=1}^N {\partial Q\over \partial \tilde{\phi}_i} {d \tilde{\phi}_i \over dr} dr, 
\end{equation}
and we could obtain $\mathcal{L}_\text{BPS}$ in terms of the effective fields and their first-derivative.


Now we proceed to find the $\Phi_i$s from $\mathcal{L}-\mathcal{L}_{BPS}=\sum_{i=1}^N \Phi_i(\vec{\phi},\partial\vec{\phi})$. As we mentioned $\Phi_i$ must be positive-semidefinite function and we restric it has to be a function of $\vec{\tilde{\phi}}$ and $\partial_r\tilde{\phi}_i$ for each $i=1,\ldots,N$. The BPS equation $\Phi_i=0$ gives solutions to $\partial_r\tilde{\phi}_i$ as follows
\begin{equation}
\partial_r\tilde{\phi}_i= \left\{F^{(1)}_i,F^{(2)}_i,...,F^{(m)}_i\right\},
\end{equation}
with $F^{(k)}_i=F^{(k)}_i(\vec{\tilde{\phi}};r)$ ($k=1,...,m$). Positive-semidefinite condition fixes $m$ to be an even number and further there must be even number of equals solutions in $\{F^{(k)}_i\}$. As an example if $m=2$ for all $i$ then $\Phi_i=0$ is a quadratic equation in $\partial_r\tilde{\phi}_i$ and so we will have $F^{(1)}_i=F^{(2)}_i$. The restriction on $\Phi_i\equiv\Phi_i(\partial_r\tilde{\phi}_i)$ forces us to rewrite the function $\mathcal{L}-\mathcal{L}_{BPS}$ into partitions $\sum_{i=1}^N\Phi_i$ explicitly. This is difficult to apply on more general forms of Lagrangian, since there exists a possibility that there are terms with $\partial\tilde{\phi}_i\partial\tilde{\phi}_j$ where $i\neq j$. Another problem is ambiguity in choosing which terms contain non-derivative of fields that should belong to which partitions $\Phi_i$.

For more general situations, the BPS equations can be obtained by procedures explained in~\cite{Atmaja:2015umo} which we describe below. On a closer look, we can consider $\mathcal{L}-\mathcal{L}_\text{BPS}=0$ as a polynomial equation of first-derivative fields. Seeing it as the polynomial equation of $\partial_r\tilde{\phi}_1$, whose maximal power is $m_1$, its roots are
\begin{equation}
\partial_r\tilde{\phi}_1= \left\{G^{(1)}_1,G^{(2)}_1,\ldots,G^{(m_1)}_1\right\},
\end{equation}
with $G^{(k)}_1=G^{(k)}_1(\vec{\tilde{\phi}},\partial_r\tilde{\phi}_2,\ldots,\partial_r\tilde{\phi}_N;r)$ and $k=1,...,m_1$. Then we have
\begin{equation}
\Phi_1\propto\left(\partial_r\tilde{\phi}_1-G^{(1)}_1\right)\left(\partial_r\tilde{\phi}_1-G^{(2)}_1\right)\ldots\left(\partial_r\tilde{\phi}_1-G^{(m_1)}_1\right).
\end{equation}
As we mentioned before here $m_1$ must be an even number and to ensure positive-definiteness at least two or more even number of roots must be equal. This will result in some constraint equations that are polynomial equations of the remaining first-derivative fields $(\partial_r\tilde{\phi}_2,\ldots,\partial_r\tilde{\phi}_N)$. Repeating the previous procedures for $\partial_r\tilde{\phi}_2$ until $\partial_r\tilde{\phi}_N$ whose $\Phi_N$ is
\begin{equation}
\Phi_N\propto\left(\partial_r\tilde{\phi}_N-G^{(1)}_N\right)\left(\partial_r\tilde{\phi}_N-G^{(2)}_N\right)...\left(\partial_r\tilde{\phi}_N-G^{(m_N)}_N\right),
\end{equation}
with $m_N$ is also an even number. Now all $G^{(k)}_N$ are only functions of $\vec{\tilde{\phi}}$ and equating some of the roots will become constraint equations that we can solve order by order for each power series of $r$. As an example let take $N=2$ and $m_1,m_2=2$. Then the constraint $G^{(1)}_1-G^{(2)}_1=0$ can be seen as a quadratic equation of $\partial_r\tilde{\phi}_2$. This give us the last constraint $G^{(1)}_2-G^{(2)}_2=0$. Since the model is valid for all $r$, we could write the constraint as $G^{(1)}_2-G^{(2)}_2=\sum_{n}a_n r^n$, where all $a_n$s are independent of $\partial_r\tilde{\phi}_1$ and $\partial_r\tilde{\phi}_2$. Then all $a_n$s need to be zero and from them we can find each $Q_i(\tilde{\phi}_i)$. Then the BPS equations for $\partial_r\tilde{\phi}_i$ can be found. 

We can see that this more general method is straightforward for any Lagrangian. This will be used throughout this paper, since we will later use some DBI-type Lagrangian that contains terms inside square root which is not easy to write the partitions explicitly. 
In \cite{Atmaja:2015umo}, with particular ansatz for the fields, writing $Q=2\pi F(f) A(a)$ is shown to be adequate for some models of vortices. Here, we show that the method is also able to do the job, at least for some known models of magnetic monopoles and dyons, using the well-known 't Hooft-Polyakov ansatz.

\section{The 't Hooft-Polyakov Monopole and Julia-Zee Dyon}
The model is described in a flat $(1+3)$-dimensional space-time whose Minkowskian metric is $\eta_{\mu\nu}=\mathrm{diag}(1,-1,-1,-1)$. The standard Lagrangian for BPS monopole, or the $SU(2)$ Yang-Mills-Higgs model, has the following form~\cite{Polyakov:1974ek,tHooft:1974kcl}
\begin{equation}\label{eq:canonicalLagrangian}
\mathcal{L}_\text{s}={1\over 2} \mathcal{D}_\mu \phi^a \mathcal{D}^\mu \phi^a -{1\over 4} F^{a\mu\nu} F^{a}_{~\mu\nu}  - V(\vert\phi\vert),
\end{equation}
with $SU(2)$ gauge group symmetry and $\phi^a$, $a=1,2,3$, is a triplet real scalar field in adjoint representation of $SU(2)$. The potensial $V$ is a function of $|\phi|=\phi^a\phi^a$ which is invariant under $SU(2)$ gauge transformations. Here we use Einstein summation convention for repeated index. The definitions of covariant derivative and field strength tensor of the $SU(2)$ Yang-Mills gauge field are as follows
\begin{subequations}
\begin{align}
\mathcal{D}_\mu \phi^a=&\partial_\mu \phi^a + e~\epsilon^{abc}A^b_\mu \phi^c,\\
F^a_{\mu\nu}=&\partial_\mu A^a_\nu - \partial_\nu A^a_\mu +e~\epsilon^{abc} A^b_\mu A^c_\nu,
\end{align}
\end{subequations}
with $e$ is the gauge coupling and $\epsilon^{abc}$ is the Levi-Civita symbol. The latin indices $(a,b,c)$ denote the ``vector components'' in the vector space of $SU(2)$ algebra with generators $T_a={1\over2}\sigma_a$, where $\sigma_a$ is the Pauli's matrix. The generators satisfy commutation relation $[T_a,T_b]=i\epsilon_{abc}T_c$ and their trace is $\mathrm{tr}(T_aT_b)={1\over2}\delta_{ab}$. With these generators, the scalar field, gauge field, adjoint covariant derivative and field strength tensor can then be re-written in a compact form, respectively, as $\phi=\phi^a T_a$, $A_\mu =A^a_\mu T_a$,
\begin{subequations}\label{eq:derivedLagrangian}
\begin{align}
\mathcal{D}_\mu \phi=&\mathcal{D}_\mu \phi^a T_a=\partial_\mu \phi -ie[A_\mu, \phi],\\
F_{\mu\nu}=&F_{\mu\nu}^aT_a=\partial_\mu A_\nu-\partial_\nu A_\mu-ie[A_\mu,A_\nu].
\end{align}
\end{subequations}
These lead to the Lagrangian
\begin{equation}
\mathcal{L}_\text{s}=\mathrm{tr} \left(\mathcal{D}_\mu \phi \mathcal{D}^\mu \phi-{1\over 2}F^{\mu\nu} F_{\mu\nu} \right)-V(|\phi|).
\end{equation}

Variying \eqref{eq:canonicalLagrangian} with respect to the scalar field and the gauge field yields
\begin{subequations}
\begin{align}
\mathcal{D}_\mu\left(\mathcal{D}^\mu\phi^b\right)=&-{\partial V\over \partial \phi^b},\label{eq:EoMscalar}\\
\mathcal{D}_\nu F^{b\mu\nu}=&e\epsilon^{bca}\phi^c\mathcal{D}^\mu\phi^a,\label{eq:EoMgauge}
\end{align}
\end{subequations}
with additional Bianchi identity
\begin{equation}
\mathcal{D}_\mu \tilde{F}^{a\mu\nu}=0,
\end{equation}
where $\tilde{F}^{a\mu\nu}={1\over2}\epsilon^{\mu\nu\kappa\lambda}{F}^a_{\kappa\lambda}$.
Through out this paper, we will consider only static configurations. The difference between monopole and dyon is whether $A^a_0$ is zero or non-zero, respectively. For monopole, the Bianchi identity becomes 
\begin{equation}
\mathcal{D}_iB^a_i=0.\label{Bianchi}
\end{equation}
Here $B^a_i={1\over 2}\epsilon_{ijk}F_{jk}$ and $i,j,k=1,2,3$ are the spatial indices.
For dyon, $A^a_0\neq0$, there are additional equations of motion for ``electric'' part since the Gauss law is non-trivial,
\begin{equation}
\mathcal{D}_i E^b_i=-e~\epsilon^{bca}\phi^c\mathcal{D}_0\phi^a,\label{Gauss}
\end{equation}
where $E^a_i=F^a_{0i}$. 

We could write the energy-momentum tensor $T_{\mu\nu}$  by varying the action with respect to the space-time metric. The energy density is then given by $T_{00}$ component,
\begin{align}
T_{00}={1\over2}\left(\mathcal{D}_0\phi^a\mathcal{D}_0\phi^a+\mathcal{D}_i\phi^a\mathcal{D}_i\phi^a+E^a_iE^a_i+B^a_iB^a_i\right)+V(|\phi|).
\end{align}
In \cite{Prasad:1975kr}, it is possible to obtain the exact solutions of the Euler-Lagrange equations in the BPS limit, i.e. $V=0$ but still maintaining the asymptotic boundary conditions of $\phi$, and we define a new parameter $\alpha$ such that
\begin{align}
T_{00}=&{1\over2}\left(\mathcal{D}_0\phi^a\mathcal{D}_0\phi^a+\mathcal{D}_i\phi^a\mathcal{D}_i\phi^a\sin^2\alpha+E^a_iE^a_i+\mathcal{D}_i\phi^a\mathcal{D}_i\phi^a\cos^2\alpha+B^a_iB^a_i\right)\nonumber\\
=&{1\over2}\left((\mathcal{D}_0\phi^a)^2+(\mathcal{D}_i\phi^a\sin\alpha\mp E^a_i)^2+(\mathcal{D}_i\phi^a\cos\alpha\mp B^a_i)^2\right) \pm E^a_i \mathcal{D}_i\phi^a\sin\alpha\pm B^a_i \mathcal{D}_i\phi^a\cos\alpha.\label{eq:alpha}
\end{align}
The last two terms can be converted to total derivative
\begin{subequations}
\begin{align}
E^a_i \mathcal{D}_i\phi^a=\partial_i(E^a_i\phi^a)-(\mathcal{D}_iE^a_i) \phi^a=\partial_i(E^a_i\phi^a),\\
B^a_i \mathcal{D}_i\phi^a=\partial_i(B^a_i\phi^a)-(\mathcal{D}_iB^a_i) \phi^a=\partial_i(B^a_i\phi^a),
\end{align}
\end{subequations}
after employing the Gauss law (\ref{Gauss}) and Bianchi identity (\ref{Bianchi}). They are related to the ``Abelian'' electric and magnetic fields identified in~\cite{tHooft:1974kcl}, respectively. Since the total energy is $E=\int d^3x~ T_{00}$, the total derivative terms can be identified as the electric and magnetic charges accordingly
\begin{subequations}
 \begin{align}
  \mathcal{Q}_E=\int dS^i E^a_i \phi^a,\\
  \mathcal{Q}_B=\int dS^i B^a_i \phi^a,
 \end{align}
\end{subequations}
with $dS^i$ denoting integration over the surface of a 2-sphere at $r\to\infty$.
Therefore the total energy is $E\geq\pm\left(\mathcal{Q}_E\sin\alpha+\mathcal{Q}_B\cos\alpha\right)$ since the other terms are positive semi-definite. The total energy is saturated if the BPS equations are satisfied as folows~\cite{Coleman:1976uk}
\begin{subequations}
\label{BPS dyons}
\begin{align}
\mathcal{D}_0\phi^a=0,\\
\mathcal{D}_i\phi^a\sin\alpha=&E^a_i,\\
\mathcal{D}_i\phi^a\cos\alpha=&B^a_i.
\end{align}
Solutions to these equations are called BPS dyons; they are particullary called BPS monopoles for $\alpha=0$. The energy of this BPS configuration is simply given by
\begin{equation}
E_{BPS}=\pm\left(\mathcal{Q}_E\sin\alpha+\mathcal{Q}_B\cos\alpha\right).
\end{equation}
\end{subequations}
Adding the constant $\alpha$ contained in $\sin\alpha$ and $\cos\alpha$ is somehow a bit tricky. We will show later using BPS Lagrangian method that this constant comes naturally as a consequence of idenfitiying two of the effective fields.

Employing the 't Hooft-Polyakov, together with Julia-Zee, ansatz~\cite{Polyakov:1974ek,tHooft:1974kcl,Julia:1975ff}
\begin{subequations}\label{eq:ansatz}
\begin{align}
\phi^a&= f(r) {x^a\over r},\\
A^a_0&={j(r)\over e} {x^a\over r},\\
A^a_i&={1-a(r)\over e} \epsilon^{aij} {x^j\over r^2},
\end{align}
\end{subequations}
where $x^a\equiv(x,y,z)$, and $x^i\equiv(x,y,z)$ as well, denotes the Cartesian coordinate. Notice that the Levi-Civita symbol $\epsilon^{aij}$ in (\ref{eq:ansatz}) mixes the space-index and the group-index. Substituting the ansatz (\ref{eq:ansatz}) into Lagrangian \eqref{eq:canonicalLagrangian} we can arrive at the following effective Lagrangian
\begin{align}
\mathcal{L}_\text{s}
=&- {f'^2\over 2} -\left({ af\over r}\right)^2  
+ {j'^2\over 2e^2} +\left({aj\over er}\right)^2
- \left({a'\over er}\right)^2 -{1\over 2} \left({a^2-1\over er^2}\right)^2 - V(f),
\label{eq:effcanLag}
\end{align}
where $'\equiv {\partial \over \partial r}$ otherwise it means taking derivative over the argument. As shown in the effective Lagrangian above there is no dependency over angles coordinates $\phi$ and $\theta$ despite the fact that the ansatz (\ref{eq:ansatz}) depends on $\phi$ and $\theta$. Thus we may derive the Euler-Lagrange equations from the effective Lagrangian (\ref{eq:effcanLag}) which are given by
\begin{subequations}\label{eq:EoMstd}
\begin{align}
-{1\over r^2}(r^2 f')'+{2 a^2f\over r^2}=&-V'(f),\\
-{(r^2 j')'\over er^2}+{2a^2j\over er^2}=&0,\\
{a(a^2-1)\over r^2}+a(e^2f^2-j^2)-a''=&0.
\end{align}
\end{subequations}
Later we will also consider the case for generalize Lagrangian of (\ref{eq:canonicalLagrangian}) by adding scalar-dependent couplings to the kinetic terms as follows~\cite{Casana:2012un}
\begin{equation}\label{eq:LagcanG}
\mathcal{L}_\text{G}=-{1\over 4}w(|\phi|)F^a_{\mu\nu}F^{a\mu\nu}+{1\over 2}G(|\phi|)\mathcal{D}_\mu\phi^a\mathcal{D}^\mu\phi^a-V(|\phi|).
\end{equation}
The equations of motions are now given by
\begin{subequations}
\begin{align}
\mathcal{D}_\mu\left(G~\mathcal{D}^\mu\phi^b\right)=&-{\partial V\over \partial \phi^b}+{1\over 2}{\partial G\over \partial \phi^b}\mathcal{D}_\mu\phi^a\mathcal{D}^\mu\phi^a-{1\over 4}{\partial w\over \partial \phi^b}F^a_{\mu\nu}F^{a\mu\nu},\label{EoMscalarG}\\
\mathcal{D}_\nu\left(w~F^{b\mu\nu}\right)=&e\epsilon^{bca}\phi^cG~\mathcal{D}^\mu\phi^a.\label{EoMgaugeG}
\end{align}
\end{subequations}
In \cite{Casana:2012un,Casana:2013lna}, they found BPS monopole equations and a constraint equation $G=w^{-1}$. Using our method in the following sections, we obtain the similar BPS monopole equations and constraint equation. Furthermore, we generalize it to BPS dyon equations with a more general constraint equation.

There are other forms of Lagrangian for BPS monopole and dyon which were presented in the Born-Infeld type of action by Nakamula and Shiraishi in \cite{Nakamula:2013rfa}. The Lagrangian for BPS monopole is different from the BPS dyon. The Lagrangians are defined such that the BPS equations (\ref{BPS dyons}) satisfy the Euler-Lagrange equations in the usual BPS limit. The Lagrangian for monopole and dyon are given respectively by\cite{Nakamula:2013rfa}
\begin{align}
\mathcal{L}_\text{NSm}=-b^2~\mathrm{tr}&\left(
\sqrt{1-{2\over b^2} \mathcal{D}_\mu \phi \mathcal{D}^\mu \phi}
\sqrt{1+{1\over b^2} F_{\mu\nu} F^{\mu\nu}}
-1
\right) -V(|\phi|),\label{eq:NSmonopole}\\
\mathcal{L}_\text{NSd}=-b^2~\mathrm{tr}&\left(
\left\{1-{2\over b^2} \mathcal{D}_\mu \phi \mathcal{D}^\mu \phi
+{1\over b^2} F_{\mu\nu} F^{\mu\nu}
-{1\over 4b^4}\left(F_{\mu\nu} \tilde{F}^{\mu\nu}\right)^2\right.\right.\nonumber\\
&\left.\left.+{4\over b^4}\tilde{F}_\mu^{~\nu} \tilde{F}^{\mu\lambda}\mathcal{D}_\nu\phi\mathcal{D}_\lambda\phi\right\}^{1/2}
-1\right) -V(|\phi|),\label{eq:NSdyon}
\end{align}
with $b^2$ is the Born-Infeld parameter and the potential $V$ is taken to be the same as in (\ref{eq:canonicalLagrangian}). It is apparent that, even though $E^a_i=0$, $\mathcal{L}_\text{NSd}\neq\mathcal{L}_\text{NSm}$. Using the ansatz (\ref{eq:ansatz}), both Lagrangians can be effectively writen as
\begin{align}
\mathcal{L}_\text{NSm}=-2b^2&\left(
\sqrt{1+{1\over 2b^2} \left(f'^2+{2a^2f^2\over r^2}\right)}
\sqrt{1+{1\over 2b^2} \left({2a'^2\over e^2r^2}+{(a^2-1)^2\over e^2r^4}\right)}
-1
\right) \nonumber\\&-V(f),\label{eq:effNSm}\\
\mathcal{L}_\text{NSd}
=-2b^2&\left(\left\{
1+{1\over 2b^2}\left(
f'^2+{2a^2f^2\over r^2}
+{2a'^2\over e^2r^2}+{(a^2-1)^2\over e^2r^4}
-{j'^2\over e^2}-{2a^2j^2\over e^2 r^2}\right)
\right.\right.\nonumber\\&\left.\left.
+{1\over4b^4}\left(
-\left(-{(a^2-1)j'\over e^2r^2}-{2aja'\over e^2r^2}\right)^2
+\left({(a^2-1)f'\over er^2}+{2 afa'\over er^2}\right)^2
\right)\right\}^{1/2}
\right.\nonumber\\&\left.-1\right)-V(f).\label{eq:effNSd}
\end{align}
We can see immeditely that $\mathcal{L}_\text{NSd}(j=0)\neq\mathcal{L}_\text{NSm}$. However, by assuming the BPS equations $B^a_i=\pm\mathcal{D}_i\phi^a$ is valid beforehand we would get $\mathcal{L}_\text{NSd}=\mathcal{L}_\text{NSm}$. Hence from both Lagrangians, we could obtain the same BPS equations when we turn off the ``electric'' part for monopole.

\section{BPS Equations in $SU(2)$ Yang-Mills-Higgs and Nakamula-Shiraishi Models}
Here we will show that the BPS Lagrangian method~\cite{Atmaja:2015umo} can also be used to obtain the known BPS equations for monopole and dyon in the $SU(2)$ Yang-Mills model \eqref{eq:canonicalLagrangian}, and the Nakamula-Shiraishi models, \eqref{eq:NSmonopole} and \eqref{eq:NSdyon}. To simplify our calculations, from here on we will set the gauge coupling to unity, $e=1$.

\subsection{BPS monopole and dyon in $SU(2)$ Yang-Mills-Higgs model}
Writing the ansatz \eqref{eq:ansatz} in spherical coordinates,
\begin{subequations}\label{spherical ansatz}
\begin{align}
\phi^a &\equiv f (\cos\varphi\sin\theta,\sin\varphi\sin\theta,\cos\theta),\\
A^a_0 &\equiv {j} (\cos\varphi\sin\theta,\sin\varphi\sin\theta,\cos\theta),\\
A^a_r &\equiv (0,0,0),~\\
A^a_\theta&\equiv {(1-a)}(\sin\varphi,-\cos\varphi,0),\\
A^a_\varphi &\equiv {(1-a)} \sin\theta (\cos\varphi\cos\theta,\sin\varphi\cos\theta,-\sin\theta),
\end{align}
\end{subequations}
we find that there is no explicit $r$ dependent in all fields above. Therefore we propose that the BPS energy function for the case of monopole, where $j=0$, should take the following form
\begin{equation}
Q(a,f)=4\pi F(f)A(a).\label{eq:Q}
\end{equation}
Since $\int d^3x~ \mathcal{L}_{\text{BPS}}=-\int dQ$, we have the BPS Lagrangian
\begin{equation}
\mathcal{L}_{\text{BPS}}=-{FA'(a)\over r^2}a'-{F'(f)A\over r^2}f'.\label{eq:LagBPS}
\end{equation}
Before showing our results, for convenience we define through all calculations in this article $x=f',~y=a',~ Q_a=F~A'(a),$ and $Q_f=F'(f)A$.

Employing $\mathcal{L}_\text{s}-\mathcal{L}_\text{BPS}=0$, where $\mathcal{L}_\text{s}$ is \eqref{eq:effcanLag} and $\mathcal{L}_\text{BPS}$ is \eqref{eq:LagBPS}, we can consider it as a quadratic equation of either $a'$ or $f'$. Here we show the roots of $f'$ (or $x$) first which are
\begin{equation}\label{eq:YMH monopole 1}
f'_\pm=\frac{Q_f \pm\sqrt{ Q_f^2-a^4-2 a^2 \left(f^2 r^2-1\right)-2 r^2 y (y-Q_a)-2 r^4 V-1}}{r^2}.
\end{equation}
The two roots will be equal, $f_+=f_-$, if the terms inside the square-root is zero, which later can be considered as a quadratic equation for $a'$ (or $y$) with roots
\begin{equation}\label{eq:YMH monopole 2}
a'_\pm= {1\over 2}Q_a \pm{1\over 2r}\sqrt{-2 a^4+a^2 \left(4-4 f^2 r^2\right)+Q_a^2 r^2+2 Q_f^2-4 r^4 V-2}.
\end{equation}
Again, we need the terms inside the square-root to be zero for two roots to be equal, $a_+=a_-$.  The last equation can be written in power series of $r$,
\begin{equation}
\left(2 Q_f^2-2 \left(a^2-1\right)^2\right)+ \left(Q_a^2-4 a^2 f^2\right)r^2-4 V~r^4=0,
\end{equation}
Demanding it is valid for all values of $r$, we may take $V=0$, which is just the same BPS limit in~\cite{Prasad:1975kr}. From the terms with quadratic and zero power of $r$, we obtain
\begin{align}
FA'(a)=& \pm 2 a f,\\
F'(f)A=& \pm(a^2-1),
\end{align}
which implies
\begin{equation}
FA=\pm(a^2-1)f.
\end{equation}
Inserting this into equations (\ref{eq:YMH monopole 1}) and (\ref{eq:YMH monopole 2}), we reproduce the known BPS equations for monopole,
\begin{subequations}\label{eq:knownBPS}
\begin{align}
f'=&\pm\frac{a^2-1}{r^2},\\
a'=&\pm af.
\end{align}
\end{subequations}

Now let us take $j(r)\neq0$ and consider the BPS limit, $V\to 0$. In this BPS limit, we can easly see from the effective Lagrangian (\ref{eq:effcanLag}), the Euler-Lagrange equations for both fields $f$ and $j$ are equal. Therefore it is tempted to identify $j\propto f$. Let us write it explicitly as
\begin{equation}
j(r)=\beta f(r),\label{eq:beta}
\end{equation}
where $\beta$ is a real valued constant. With this identification, we can again use \eqref{eq:Q} as the BPS energy function for dyon and hece give the same BPS Lagrangian (\ref{eq:LagBPS}). Now the only difference, from the previous monopole case, is the effective Lagrangian (\ref{eq:effcanLag}) which takes a simpler form
\begin{equation}
 \mathcal{L}_\text{s}
=- \left(1-\beta^2\right)\left({f'^2\over 2} +\left({ af\over r}\right)^2\right) 
- \left({a'\over r}\right)^2 -{1\over 2} \left({a^2-1\over r^2}\right)^2-V.
\end{equation}
Here we still keep the potential $V$ and we will show later that $V$ must be equal to zero in order to get the BPS equations using the BPS Lagrangian method.

Applying \eqref{eq:beta} and solving $\mathcal{L}_\text{s}-\mathcal{L}_\text{BPS}=0$ as quadratic equation for $f'$ (or $x$) give us two roots
\begin{equation}
f'_\pm=\frac{Q_f\pm\sqrt{\text{DD}}}{\left(1-\beta ^2\right) r^2},
\end{equation}
with
\begin{equation}
\text{DD}=\left(\beta ^2-1\right) \left(a^4-2 a^2 \left(\left(\beta ^2-1\right) f^2 r^2+1\right)+2 r^2 y (y-Q_a)+2 r^4 V+1\right)+Q_f^2.
\end{equation}
Next, requiring $f'_+=f'_-$, we obtain
\begin{equation}
a'_\pm=\frac{1}{2} \left(Q_a\pm\sqrt{\frac{\text{DDD}}{\left(\beta ^2-1\right) r^2}}\right),
\end{equation}
where we arrange $\text {DDD}$ in power series of $r$, i.e.
\begin{align}
\text {DDD}=2\left(\left(1-a^2\right)^2\left(1-\beta^2\right)-Q_f^2\right)+\left(1-\beta ^2\right)\left(4 a^2 \left(1-\beta ^2\right) f^2-Q_a^2\right) r^2+4 V\left(1-\beta ^2\right)r^4.
\end{align}
Again, $a'_-=a'_+$, we get $DDD=0$. Solving the last equation, which must be valid for all values of $r$, we conclude $V=0$ from $r^4$-terms, for non-trivial solution, and from the remaining terms we have
\begin{subequations}
\begin{align}
FA'(a)=& \pm 2 af \sqrt{1-\beta ^2},\\
F'(f)A=& \pm \left(a^2-1\right) \sqrt{1-\beta ^2},
\end{align}
\end{subequations}
which give us
\begin{equation}
FA=\pm \left(a^2-1\right)f \sqrt{1-\beta ^2}.
\end{equation}
The BPS equations are then
\begin{subequations}\label{eq:knownBPSdyon}
\begin{align}
f'\sqrt{1-\beta ^2}=&\pm\frac{a^2-1}{r^2},\\
a'=&\pm af\sqrt{1-\beta ^2}.
\end{align}
\end{subequations}
Since $f'$ and $a'$ are real-valued, $\beta$ should take values $|\beta|< 1$. They become the BPS equations for monopole \eqref{eq:knownBPS} when we set $\beta=0$. We can see that this constant is analogous to the constant $\alpha$, or precisely $\beta=-\sin\alpha$, in \eqref{eq:alpha}, see~\cite{weinberg2012classical} for detail. Substituting $\beta=-\sin\alpha$ into equations \eqref{eq:knownBPSdyon}, we get the same BPS equations as in \cite{Prasad:1975kr,Coleman:1976uk}. Here we can see the constant $\beta$ is naturally bounded as required by the BPS equations (\ref{eq:knownBPSdyon}).

\subsection{BPS monopole and dyon in Nakamula-Shiraishi model}
In this subsection we will show that the Lagrangians (\ref{eq:NSmonopole}) and (\ref{eq:NSdyon}) of Nakamula-Shiraishi model do indeed posses the BPS equations \eqref{eq:knownBPS} (and \eqref{eq:knownBPSdyon}) respectively after employing the BPS Lagrangian method. Substituting \eqref{eq:effNSm} and \eqref{eq:Q} into $\mathcal{L}_\text{NSm}-\mathcal{L}_\text{BPS}=0$ and following the same procedures as the previous subsection give us the roots of $a'$,
\begin{align}
a'_\pm=\frac{Q_a r^4 \left(2 b^2-V\right)+Q_a Q_f r^2 x\pm\sqrt{r^2 \left(2 a^2 f^2+r^2 \left(2 b^2+x^2\right)\right) \text{DD}}}{r^2 \left(4 a^2 f^2+2 r^2 \left(2 b^2+x^2\right)-Q_a^2\right)},
\end{align}
where 
\begin{align}
\text{DD}=&-4 a^6 f^2+a^4 \left(-2 r^2 \left(2 b^2+x^2\right)+8 f^2+Q_a^2\right)\nonumber\\
&-2 a^2 \left(f^2 \left(4 b^2 r^4+2\right)-2 r^2 \left(2 b^2+x^2\right)+Q_a^2\right)\nonumber\\
&+2 b^2 r^4 \left(Q_a^2+4 Q_f x-2 r^2 \left(2 V+x^2\right)\right)-4 b^2 r^2+Q_a^2\nonumber\\
&+2 r^2 \left(-Q_f x+r^2 V+x\right) \left(r^2 V-(Q_f+1) x\right).
\end{align}
Solving $\text{DD}=0$ give us
\begin{align}
f'_\pm=\frac{2 Q_f r^4 \left(2 b^2-V\right)\pm \sqrt{-2r^2 \left(\left(a^2-1\right)^2+2 b^2 r^4\right) \text{DDD}}}{2 r^2 \left(a^4-2 a^2+2 b^2 r^4-Q_f^2+1\right)},
\end{align}
where
\begin{align}\label{4b^2}
\text{DDD}=&2 b^2 r^4 \left(4 a^2 f^2-Q_a^2\right)+4 b^2 r^2 \left(\left(a^2-1\right)^2-Q_f^2\right)\nonumber\\
&+\left(\left(a^2-1\right)^2-Q_f^2\right) \left(4 a^2 f^2-Q_a^2\right)+2 r^6 V \left(4 b^2-V\right).
\end{align}
Then the last equation $\text{DDD}=0$ give us $V=0$, or $V=4b^2$,
\begin{subequations}
\begin{align}
FA'(a)=&\pm 2af,\\
F'(f)A=&\pm (a^2-1),
\end{align}
\end{subequations}
which again give us
\begin{equation}
FA=\pm(a^2-1)f,
\end{equation}
and thus we have $a'$ and $f'$, with $V=0$, 
\begin{subequations}
\begin{align}
f'=&\pm\frac{a^2-1}{r^2},\\
a'=&\pm af.
\end{align}
\end{subequations}
the same BPS equations \eqref{eq:knownBPS} for monopole. The other choice of potential $V=4b^2$ will result the same BPS equations with opposite sign relative to the BPS equations of $V=0$,
\begin{subequations}
\begin{align}
f'=&\mp\frac{a^2-1}{r^2},\\
a'=&\mp af.
\end{align}
\end{subequations}

For dyon, using the same identification \eqref{eq:beta}, we have the effective Lagrangian \eqref{eq:effNSd} shortened to
\begin{align}
\mathcal{L}_\text{NSd}=-2b^2&\left(\left\{
1+{1-\beta^2\over 2b^2}\left(
f'^2+{2a^2f^2\over r^2}\right)
+{1\over 2b^2}\left({2a'^2\over r^2}+{(a^2-1)^2\over r^4}\right)
\right.\right.\nonumber\\&\left.\left.
+{1-\beta^2\over4b^4}\left({(a^2-1)f'\over r^2}+{2afa'\over r^2}\right)^2
\right\}^{1/2}
-1\right)-V.
\end{align}
Equating the above effective Lagrangian with $\mathcal{L}_\text{BPS}$, using the same BPS energy density \eqref{eq:Q}, and solving this for $a'$ give us
\begin{align}
a'_\pm=\frac{-2 a^3 \beta ^2 f x+2 a^3 f x+2 a \beta ^2 f x-2 a f x-2 b^2 Q_a r^2-Q_a Q_f x+Q_a r^2 V\pm\sqrt{\text{DD}}}{4 a^2 \left(\beta ^2-1\right) f^2-4 b^2 r^2+Q_a^2},
\end{align}
where
\begin{align}
\text{DD}=&\left(2 a \left(a^2-1\right) \left(\beta ^2-1\right) f x+2 b^2 Q_a r^2+Q_a \left(Q_f x-r^2 V\right)\right)^2\nonumber\\
+&\left(4 a^2 \left(\beta ^2-1\right) f^2-4 b^2 r^2+Q_a^2\right) \times\nonumber\\
&\left[2 b^2 \left(a^4-2 a^2 \left(\left(\beta ^2-1\right) f^2 r^2+1\right)-2 Q_f r^2 x+r^4 \left(2 V-\beta ^2 x^2+x^2\right)+1\right)
\right.\nonumber\\
&\left.-x^2 \left(a^4 \left(\beta ^2-1\right)-2 a^2 \left(\beta ^2-1\right)+\beta ^2+Q_f^2-1\right)+2 Q_f r^2 V x-r^4 V^2\right].
\end{align}
Solving $\text{DD}=0$ for $f'$ give us
\begin{align}
f'_\pm=&\frac{\text{K}\pm\sqrt{2}\sqrt{\text{M}~\text{DDD}}}{\text{L}}
\end{align}
where
\begin{align}
\text{K}=&2 a^3 \left(\beta ^2-1\right) f Q_a r^2 \left(2 b^2-V\right)-4 a^2 \left(\beta ^2-1\right) f^2 Q_f r^2 \left(2 b^2-V\right)
\nonumber\\&
-2 a \left(\beta ^2-1\right) f Q_a r^2 \left(2 b^2-V\right)+8 b^4 Q_f r^4-4 b^2 Q_f r^4 V,\\
\text{L}=&-4 a \left(a^2-1\right) \left(\beta ^2-1\right) f Q_a Q_f\nonumber\\
&-4 \left[a^4 b^2 \left(\beta ^2-1\right) r^2-a^2 \left(\beta ^2-1\right) \left(f^2 \left(2 b^2 \left(\beta ^2-1\right) r^4+Q_f^2\right)+2 b^2 r^2\right)
\right.\nonumber\\&\left.
+b^2 r^2 \left(\left(\beta ^2-1\right) \left(2 b^2 r^4+1\right)+Q_f^2\right)\right]
\nonumber\\
&+\left(\beta ^2-1\right) Q_a^2 \left(a^4-2 a^2+2 b^2 r^4+1\right),\\
\text{M}=&-b^2 \left(a^4-2 a^2 \left(\left(\beta ^2-1\right) f^2 r^2+1\right)+2 b^2 r^4+1\right) \left(4 a^2 \left(\beta ^2-1\right) f^2-4 b^2 r^2+Q_a^2\right),\\
\text{DDD}=&-2 r^4 \left(b^2 \left(\beta ^2-1\right) \left(4 a^2 \left(\beta ^2-1\right) f^2+Q_a^2\right)\right)-\left(\beta ^2-1\right) \left(\left(a^2-1\right) Q_a-2 a f Q_f\right)^2\nonumber\\
&+4 b^2 r^2 \left(a^4 \left(\beta ^2-1\right)-2 a^2 \left(\beta ^2-1\right)+\beta ^2+Q_f^2-1\right)+2 \left(\beta ^2-1\right) r^6 V \left(4 b^2-V\right).
\end{align}
We may set $M=0$, but this will imply $b^2=0$ which is not what we want. Requiring $\text{DDD}=0$  valid for all values of $r$, the terms with $r^6$ give us $V=0$ or $V=4b^2$. The terms with $r^0$ imply
\begin{equation}
FA'(a)= \frac{2 a f F'(f)A}{a^2-1}.
\end{equation}
This is indeed solved by the remaining terms which imply
\begin{align}
FA'(a)=&\pm 2af \sqrt{1-\beta ^2},\\
F'(f)A=&\pm \left(a^2-1\right) \sqrt{1-\beta ^2}.
\end{align}
This again give us
\begin{equation}
FA=\pm \left(a^2-1\right)f \sqrt{1-\beta ^2},
\end{equation}
hence, for $V=0$,
\begin{subequations}
\begin{align}
f'=&\pm\frac{a^2-1}{\sqrt{1-\beta ^2} r^2},\\
a'=&\pm af\sqrt{1-\beta ^2},
\end{align}
\end{subequations}
the same BPS equations \eqref{eq:knownBPSdyon} for dyon. Similar to the monopole case choosing $V=4b^2$ will switch the sign in the BPS equations. It is apparent that it the limit of $\beta\to0$, the BPS equations for dyon becomes the ones for monopole. This indicates that in the BPS limit and $\beta\to0$, $\mathcal{L}_\text{NSd}\to\mathcal{L}_\text{NSm}$, since in general, even though in the limit of $\beta\to0$, $\mathcal{L}_\text{NSd}\slashed{\to}\mathcal{L}_\text{NSm}$.

Now we know that the method works. In the next sections, we use it in some generalized Lagrangian whose BPS equations, for monopole or dyon, may or may not be known.

\section{BPS Equations in Generalized $SU(2)$ Yang-Mills-Higgs Model}
In this section, we use the Lagrangian \eqref{eq:LagcanG} whose its effective Lagrangian is given by
\begin{equation}\label{eq:gencanLag}
\mathcal{L}_\text{G}=-G \left(\frac{f'^2}{2}+\frac{a^2 f^2}{r^2}\right)+w \left(\frac{j'^2}{2}+\frac{a^2 j^2}{r^2}\right)-w \left(\frac{a'^2}{r^2}+\frac{\left(a^2-1\right)^2}{2 r^4}\right)-V.
\end{equation}
We will see later it turns out that $G$ and $w$ are related to each other by some constraint equations.

\subsection{BPS monopole case}
In this case, the BPS equations are already known \cite{Casana:2012un,Casana:2013lna}. Setting $j=0$ and employing $\mathcal{L}_\text{G}-\mathcal{L}_\text{BPS}=0$ we get
\begin{equation}
f'_\pm=\frac{Q_f r^2\pm\sqrt{-r^4 \left(G \left(a^4 w+2 a^2 \left(f^2 G r^2-w\right)+2 r^2 y (w y-Q_a)+2 r^4 V+w\right)-Q_f^2\right)}}{G r^4},
\end{equation}
and from $f'_+=f'_-$ we have the roots of $a'$ (or $y$)
\begin{align}
a'_\pm=&\frac{G Q_a r^2-\sqrt{G r^2 \left\{G \left(Q_a^2 r^2-2 w \left(2 r^2 \left(a^2 f^2 G+r^2 V\right)+\left(a^2-1\right)^2 w\right)\right)+2 Q_f^2 w\right\}}}{2 G r^2 w}.
\end{align}
The terms inside the curly bracket in the square root must be zero in which after rearranging in power series of $r$
\begin{equation}
2 w \left(Q_f^2-\left(a^2-1\right)^2 G w\right)+G r^2 \left(Q_a^2-4 a^2 f^2 G w\right)-4 r^4 (G V w)=0,
\end{equation}
we obtain $V=0$,
\begin{align}
FA'(a)=&\pm 2af \sqrt{Gw},\\
F'(f)A=&\pm \left(a^2-1\right) \sqrt{Gw}.
\end{align}
These imply
\begin{equation}
{\partial\over \partial f}\left( f \sqrt{G w}\right)=\sqrt{G w}
\end{equation}
, and hence
\begin{equation}
w={c\over G},\label{monopole constraint}
\end{equation}
where $c$ is a positive constant. The BPS equations are given by
\begin{subequations}\label{eq:GenBPSm}
\begin{align}
f'=&\pm\frac{\left(a^2-1\right)}{r^2}\sqrt{w\over G},\\
a'=&\pm a f\sqrt{G\over w},
\end{align}
\end{subequations}
with a constraint equation $w~G=c$, where $c$ is a positive constant. This constant can be fixed to one, $c=1$, by recalling that in the corresponding non-generalized version, in which $G=w=1$, we should get back the same BPS equations of (\ref{eq:knownBPS}).

\subsection{BPS dyon case}
As previously setting $j=\beta f$ and employing $\mathcal{L}_\text{G}-\mathcal{L}_\text{BPS}=0$ we get
\begin{equation}
f'_\pm=\frac{Q_f r^2\pm\sqrt{r^4 \text{DD}}}{r^4 \left(G-\beta ^2 w\right)},
\end{equation}
with
\begin{equation}
\text{DD}=Q_f^2-\left(G-\beta ^2 w\right) \left(a^4 w+2 a^2 \left(f^2 r^2 \left(G-\beta ^2 w\right)-w\right)+2 r^2 y (w y-Q_a)+2 r^4 V+w\right),
\end{equation}
and from $\text{DD}=0$ we have the roots of $a'$
\begin{align}
a'_\pm=&{Q_a\pm \sqrt{\text{DDD}\over r^2 \left(G-\beta ^2 w\right)}\over 2w},
\end{align}
where 
\begin{align}
\text{DDD}=&r^2 \left(G-\beta ^2 w\right) \left(4 a^2 f^2 w \left(\beta ^2 w-G\right)+Q_a^2\right)\nonumber\\
&+2 w \left(\left(a^2-1\right)^2 w \left(\beta ^2 w-G\right)+Q_f^2\right)+4 r^4 V w \left(\beta ^2 w-G\right).
\end{align}
Requiring $\text{DDD}=0$ we obtain $V=0$,
\begin{align}
FA'(a)=&\pm 2 a f \sqrt{w\left(G-\beta ^2 w\right)},\\
F'(f)A=&\pm \left(a^2-1\right) \sqrt{w\left(G-\beta ^2 w\right)} .
\end{align}
Similar to the monopole case these imply
\begin{equation}
{w(G-\beta ^2 w)}=c,\label{dyon constraint}
\end{equation}
where $c$ is a positive constant and it can also be fixed to $c=1-\beta^2$ demanding that at $G=w=1$ we should get the same BPS equations (\ref{eq:knownBPSdyon}). At $\beta\to0$, we get back the constraint equation (\ref{monopole constraint}) for monopole case.
These give us the BPS equations
\begin{subequations}\label{eq:GenBPSd}
\begin{align}
f'=&\pm\frac{\left(a^2-1\right)}{r^2}\sqrt{w\over G-\beta ^2 w},\\
a'=&\pm a f \sqrt{G-\beta ^2 w\over w},
\end{align}
\end{subequations}
in which at $\beta\to0$ we again get back the BPS equations for monopole case \eqref{eq:GenBPSm}.

\section{BPS Equations in Generalized Nakamula-Shiraishi Model}
Here we present the generalized version of the Nakamula-Shiraishi models (\ref{eq:NSmonopole}) and (\ref{eq:NSdyon}) for both monopole and dyon respectively.
\subsection{BPS monopole case}
For a generalized version of (\ref{eq:NSmonopole}) is defined by
\begin{align}
\label{NSMonopole}
\mathcal{L}_\text{NSmG}=&-b^2\mathrm{tr}\left(
\sqrt{1-{2\over b^2}G(|\phi|) \mathcal{D}_\mu \phi \mathcal{D}^\mu \phi}
\sqrt{1+{1\over b^2}w(|\phi|) \mathcal{F}_{\mu\nu} \mathcal{F}^{\mu\nu}}
-1
\right) -V(|\phi|),
\end{align}
where after inserting the ansatz, we write its effective Lagrangian as
\begin{align}
\label{NSMonopoleEff}
\mathcal{L}_\text{NSmG}=&-2b^2\left(
\sqrt{1+{G\over 2b^2} \left(f'^2+{2a^2f^2\over r^2}\right)}
\sqrt{1+{w\over 2b^2} \left({2a'^2\over r^2}+{(a^2-1)^2\over r^4}\right)}
-1\right) -V.
\end{align}
Using the similar BPS Lagrangian (\ref{eq:LagBPS}), we solve $\mathcal{L}_\text{NSmG}-\mathcal{L}_{\text{BPS}}=0$ as a quadratic equation of $a'$ (or $y$) first as such the roots are given by
\begin{align}
a'=\frac{Q_a r^4 \left(2 b^2-V\right)+Q_a Q_f r^2 x\pm\sqrt{r^2 \left(2 a^2 f^2 G+r^2 \left(2 b^2+G x^2\right)\right) \text {DD}}}{r^2 \left(2 w \left(2 a^2 f^2 G+r^2 \left(2 b^2+G x^2\right)\right)-Q_a^2\right)},
\end{align}
with
\begin{align}
\text {DD}=&w \left(\left(a^2-1\right)^2 \left(Q_a^2-2 G w \left(2 a^2 f^2+r^2 x^2\right)\right)+2 \left(r^3 V-Q_f r x\right)^2\right)
\nonumber\\
&+2 b^2 r^2 \left(Q_a^2 r^2-2 w \left(a^4 w+2 a^2 \left(f^2 G r^2-w\right)+r^4 \left(G x^2+2 V\right)-2 Q_f r^2 x+w\right)\right).
\end{align}
Taking $\text {DD}=0$, we obtain the roots for $f'$,
\begin{align}
f'=\frac{2Q_f r^4 w \left(2 b^2-V\right)\pm \sqrt{2r^2 w \left(\left(a^2-1\right)^2 w+2 b^2 r^4\right) \text {DDD}}}{2 r^2 w \left(\left(a^2-1\right)^2 G w+2 b^2 G r^4-Q_f^2\right)},
\end{align}
with
\begin{align}
\text {DDD}=&2 b^2 G r^4 \left(Q_a^2-4 a^2 f^2 G w\right)+4 b^2 r^2 w \left(Q_f^2-\left(a^2-1\right)^2 G w\right)\nonumber\\
&+\left(Q_f^2-\left(a^2-1\right)^2 G w\right) \left(4 a^2 f^2 G w-Q_a^2\right)+2 G r^6 V w \left(V-4 b^2\right).
\end{align}
Requiring $\text {DDD}=0$, we obtain from the terms with $r^6$ that $V=0$ or $V=4b^2$. The remaining terms are also zero if $Q_a=\pm 2 af \sqrt{G w}$ and $Q_f=\pm\left(a^2-1\right) \sqrt{G w}$. These again imply
\begin{equation}
G={1\over w},
\end{equation}
which is equal to the constraint equation (\ref{monopole constraint}) for monopole in Generalized $SU(2)$ Yang-Mills-Higgs model. Then the BPS equations, with $V=0$, are
\begin{subequations}
\begin{align}
f'=&\pm\frac{\left(a^2-1\right)}{r^2} \sqrt{w\over G},\\
a'=&\pm a f \sqrt{G\over w},
\end{align}
\end{subequations}
which are equal to BPS equations \eqref{eq:GenBPSm} for monopole in the Generalized $SU(2)$ Yang-Mills-Higgs model.

\subsection{BPS dyon case}
The generalization of Lagrangian (\ref{eq:NSdyon}) is defined as
\begin{align}
\label{NSDyon}
\mathcal{L}_\text{NSdG}=-b^2\mathrm{tr}&\left(
\left\{1-{2\over b^2}G(|\phi|) \mathcal{D}_\mu \phi \mathcal{D}^\mu \phi
+{1\over b^2}w(|\phi|) \mathcal{F}_{\mu\nu} \mathcal{F}^{\mu\nu}
-{1\over 4b^4}G_1(|\phi|) \left(\mathcal{F}_{\mu\nu} \tilde{\mathcal{F}}^{\mu\nu}\right)^2\right.\right.\nonumber\\
&\left.\left.+{4\over b^4}G_2(|\phi|) \tilde{\mathcal{F}}_\mu^{~\nu} \tilde{\mathcal{F}}^{\mu\lambda}\mathcal{D}_\nu\phi\mathcal{D}_\lambda\phi\right\}^{1/2}
-1\right) -V(|\phi|).
\end{align}
Employing the relation $j(f)=\beta f$, its effective Lagrangian is
\begin{align}
\label{NSDyonEff}
\mathcal{L}_\text{NSdG}=-2b^2&\left(\left\{
1+{G- w\beta^2\over 2b^2}\left(
f'^2+{2a^2f^2\over r^2}\right)
+{w\over 2b^2}\left({2a'^2\over r^2}+{(a^2-1)^2\over r^4}\right)
\right.\right.\nonumber\\&\left.\left.
+{G_2-G_1\beta^2\over4b^4}\left({(a^2-1)f'\over r^2}+{2afa'\over r^2}\right)^2
\right\}^{1/2}
-1\right)-V.
\end{align}
Employing $\mathcal{L}_\text{NSdG}-\mathcal{L}_{\text{BPS}}=0$, with the same BPS Lagrangian (\ref{eq:LagBPS}), and solving it as a quadratic equation of $a'$ first we get
\begin{align}
a'=\frac{2 a^3 \beta ^2 f G_1 x-2 a^3 f G_2 x-2 a \beta ^2 f G_1 x+2 a f G_2 x+2 b^2 Q_a r^2+Q_a Q_f x-Q_a r^2 V \pm\frac{1}{2} \sqrt{\text{DD}}}{4 a^2 f^2 \left(G_2-\beta ^2 G_1\right)+4 b^2 w r^2-Q_a^2},
\end{align}
where
\begin{align}
\text{DD}=&\left(-4 a \left(a^2-1\right) f x \left(G_2-\beta ^2 G_1\right)+4 b^2 Q_a r^2+Q_a \left(2 Q_f x-2 r^2 V\right)\right)^2 \nonumber\\
&-4 \left(-4 a^2 f^2 \left(G_2-\beta ^2 G_1\right)-4 b^2 w r^2+Q_a^2\right)\text{H},\\
\text{H}=&x^2 \left(\left(a^2-1\right)^2 \beta ^2 G_1-\left(a^2-1\right)^2 G_2+Q_f^2\right)-2 b^2 \text{J}-2 Q_f r^2 V x+r^4 V^2,\\
\text{J}=&r^2 \left(2 a^2 f^2 G-2Q_f x+r^2 \left(2 V+G x^2\right)\right)+w \left(a^4-2 a^2 \left(\beta ^2 f^2 r^2+1\right)-\beta ^2 r^4 x^2+1\right).
\end{align}
Then from $\text {DD}=0$ we have a quadratic equation of $f'$ whose roots are
\begin{align}
f'=\frac{\text {K}\pm {1\over 2}\sqrt{\text {DDD}}}{\text {L}},
\end{align}
where
\begin{align}
\text {K}=&-4 a^3 b^2 \beta ^2 f G_1 Q_a r^2+4 a^3 b^2 f G_2 Q_a r^2+2 a^3 \beta ^2 f G_1 Q_a r^2 V-2 a^3 f G_2 Q_a r^2 V\nonumber\\
&+8 a^2 b^2 \beta ^2 f^2 G_1 Q_f r^2-8 a^2 b^2 f^2 G_2 Q_f r^2-4 a^2 \beta ^2 f^2 G_1 Q_f r^2 V+4 a^2 f^2 G_2 Q_f r^2 V\nonumber\\
&+4 a b^2 \beta ^2 f G_1 Q_a r^2-4 a b^2 f G_2 Q_a r^2-2 a \beta ^2 f G_1 Q_a r^2 V+2 a f G_2 Q_a r^2 V\nonumber\\
&-8 b^4 w Q_f r^4+4 b^2 w Q_f r^4 V,\\
\text {L}=&-\beta ^2 G_1 \left(a^2 (-Q_a)+2 a f Q_f+Q_a\right)^2+2 b^2 r^2 \text {M}+G_2 ~\text {N}+8 b^4 w r^6 \left(\beta ^2 w-G\right),
\end{align}
where in L we define M and N as
\begin{align}
\text {M}=&r^2 G \left(4 a^2 \beta ^2 f^2 G_1+Q_a^2\right)+w \left(2 Q_f^2-\beta ^2 \left(Q_a^2 r^2-2 G_1 \left(a^4-2 a^2 \left(\beta ^2 f^2 r^2+1\right)+1\right)\right)\right),\\
\text {N}=&-4 a^4 b^2 w r^2+4 a^2 \left(f^2 \left(Q_f^2-2 b^2 r^4 \left(G-\beta ^2 w\right)\right)+2 b^2 w r^2\right)-4 \left(a^2-1\right) a f Q_a Q_f\nonumber\\
&+\left(a^2-1\right)^2 Q_a^2-4 b^2 w r^2,
\end{align}
and
\begin{subequations}\label{eq:complicatedDDD}
\begin{align}
\text {DDD}=T_0-8T_2r^2+8T_4r^4+16b^2T_6r^6+T_8r^8-32T_{10}r^{10}+T_{12}r^{12},
\end{align}
where
\begin{align}
T_0=&8 \left(a^2-1\right)^2 b^2 w \left(G_2-\beta ^2 G_1\right) \left(\left(a^2-1\right) Q_a-2 a f Q_f\right)^2 \left(4 a^2 f^2 \left(G_2-\beta ^2 G_1\right)-Q_a^2\right),
\end{align}
\begin{align}
T_2=&-4 \left(a^2-1\right)^2 b^4 w^2 \left(\left(a^2-1\right)^2 \beta ^2 G_1-\left(a^2-1\right)^2 G_2+Q_f^2\right) \left(4 a^2 f^2 \left(G_2-\beta ^2 G_1\right)-Q_a^2\right)\nonumber\\
&-b^2 \left(G_2-\beta ^2 G_1\right) \left(\left(a^2-1\right) Q_a-2 a f Q_f\right)^2 \times\nonumber\\
&\left(4 \left(a^2-1\right)^2 b^2 w^2-2 a^2 f^2 \left(G-\beta ^2 w\right) \left(Q_a^2-4 a^2 f^2 \left(G_2-\beta ^2 G_1\right)\right)\right),
\end{align}
\begin{align}
T_4=&2 \left(a^2-1\right)^2 b^4 w \left(\beta ^2 w-G\right) \left(Q_a^2-4 a^2 f^2 \left(G_2-\beta ^2 G_1\right)\right)^2\nonumber\\
&+2 a^2 f^2 \left(V-2 b^2\right)^2 \left(G_2-\beta ^2 G_1\right)^2 \left(\left(a^2-1\right) Q_a-2 a f Q_f\right)^2\nonumber\\
&+4 b^4 w \left(\left(a^2-1\right)^2 \beta ^2 G_1-\left(a^2-1\right)^2 G_2+Q_f^2\right)\times\nonumber\\ 
&\left(4 \left(a^2-1\right)^2 b^2 w^2-2 a^2 f^2 \left(G-\beta ^2 w\right) \left(Q_a^2-4 a^2 f^2 \left(G_2-\beta ^2 G_1\right)\right)\right)\nonumber\\
&-2 \left(G_2-\beta ^2 G_1\right) \left(\left(a^2-1\right) Q_a-2 a f Q_f\right)^2 \times\nonumber\\
&\left(b^4 \left(4 a^2 f^2 w \left(\beta ^2 w-G\right)+Q_a^2\right)-4 a^2 b^2 f^2 V \left(G_2-\beta ^2 G_1\right)+a^2 f^2 V^2 \left(G_2-\beta ^2 G_1\right)\right),
\end{align}
\begin{align}
T_6=&-4 \left(a^2-1\right)^2 b^4 w^2 \left(G-\beta ^2 w\right) \left(4 a^2 f^2 \left(G_2-\beta ^2 G_1\right)-Q_a^2\right)\nonumber\\
&+b^2 \left(G-\beta ^2 w\right) \left(Q_a^2-4 a^2 f^2 \left(G_2-\beta ^2 G_1\right)\right) \times\nonumber\\
&\left(4 \left(a^2-1\right)^2 b^2 w^2-2 a^2 f^2 \left(G-\beta ^2 w\right) \left(Q_a^2-4 a^2 f^2 \left(G_2-\beta ^2 G_1\right)\right)\right)\nonumber\\
&-4 a f w Q_f \left(V-2 b^2\right)^2 \left(G_2-\beta ^2 G_1\right) \left(\left(a^2-1\right) Q_a-2 a f Q_f\right)\nonumber\\
&+w V \left(4 b^2-V\right) \left(G_2-\beta ^2 G_1\right) \left(\left(a^2-1\right) Q_a-2 a f Q_f\right)^2\nonumber\\
&+4 w \left(\left(a^2-1\right)^2 \beta ^2 (-G_1)+\left(a^2-1\right)^2 G_2-Q_f^2\right) \times\nonumber\\
&\left(b^4 \left(4 a^2 f^2 w \left(\beta ^2 w-G\right)+Q_a^2\right)-4 a^2 b^2 f^2 V \left(G_2-\beta ^2 G_1\right)+a^2 f^2 V^2 \left(G_2-\beta ^2 G_1\right)\right),
\end{align}
\begin{align}
T_8=&64 b^4 w^2 Q_f^2 \left(V-2 b^2\right)^2-8 \left(8 b^6 w \left(G-\beta ^2 w\right) \left(4 \left(a^2-1\right)^2 b^2 w^2
\right.\right.\nonumber\\&\left.\left.
-2 a^2 f^2 \left(G-\beta ^2 w\right) \left(Q_a^2-4 a^2 f^2 \left(G_2-\beta ^2 G_1\right)\right)\right)
\right.\nonumber\\&\left.
+4 b^2 \left(G-\beta ^2 w\right) \left(Q_a^2-4 a^2 f^2 \left(G_2-\beta ^2 G_1\right)\right) \left(b^4 \left(4 a^2 f^2 w \left(\beta ^2 w-G\right)+Q_a^2\right)
\right.\right.\nonumber\\&\left.\left.
-4 a^2 b^2 f^2 V \left(G_2-\beta ^2 G_1\right)+a^2 f^2 V^2 \left(G_2-\beta ^2 G_1\right)\right)
\right.\nonumber\\&\left.
+8 b^4 w^2 V \left(4 b^2-V\right) \left(\left(a^2-1\right)^2 \beta ^2 (-G_1)+\left(a^2-1\right)^2 G_2-Q_f^2\right)\right),
\end{align}
\begin{align}
T_{10}=&b^4 w \left(\beta ^2 w-G\right) \left(4 b^4 \left(4 a^2 f^2 w \left(\beta ^2 w-G\right)+Q_a^2\right)+4 b^2 V \left(Q_a^2-8 a^2 f^2 \left(G_2-\beta ^2 G_1\right)\right)
\right.\nonumber\\&\left.
+V^2 \left(8 a^2 f^2 \left(G_2-\beta ^2 G_1\right)-Q_a^2\right)\right),
\end{align}
\begin{align}
T_{12}=128 b^6 w^2 V \left(4 b^2-V\right) \left(\beta ^2 w-G\right).
\end{align}
\end{subequations}
in order to find $V,Q_f,Q_a,G_1$ and $G_2$, we have to solve equation $\text {DDD}=0$. Since the model is valid for all $r$, then each $T_0$ until $T_{12}$ must be equal to zero. From $T_{12}=0$ we need either $V=0$ or $V=4b^2$. This verify that the BPS limit is indeed needed to obtain the BPS equations. Putting $V=0$ into $\text {DDD}$ we simplify a little the equations \eqref{eq:complicatedDDD} above. From $T_0=0$ we have 
\begin{equation}
Q_f= \frac{\left(a^2-1\right) Q_a}{2 a f},
\end{equation}
which we input into $\text {DDD}$ again. From $T_2=0$ we obtain $Q_a= \pm 2 a f \sqrt{G_2-\beta ^2 G_1}$. Now we will input each into twos separate cases.
\begin{enumerate}
\item Setting $Q_a= - 2 a f \sqrt{G_2-\beta ^2 G_1}$, only $T_8$ and $T_{10}$ are not vanished. Both can vanish if $\beta ^2 (w^2- G_1)+(G_2-w G)=0$ hence we have $G_2-\beta ^2G_1=wG-w^2\beta^2$.
\item Setting $Q_a= 2 a f \sqrt{G_2-\beta ^2 G_1}$, we also arrive at the same destination.
\end{enumerate}
From these steps, we obtain that 
\begin{align}
FA'(a)=&\pm2 a f \sqrt{w\left(G-\beta ^2 w\right)},\\
F'(f)A=&\pm(a^2-1) \sqrt{w\left(G-\beta ^2 w\right)},
\end{align}
which again imply that 
\begin{equation}
 w\left(G-\beta ^2 w\right)=1-\beta^2,
\end{equation}
which is equal to the constraint equation (\ref{dyon constraint}) for dyon in the Generalized $SU(2)$ Yang-Mills-Higgs model. Substituting everything, we obtain the BPS equations, with $V=0$,
\begin{align}
f'&=\pm\frac{\left(a^2-1\right)}{r^2}\sqrt{w\over G-\beta ^2 w},\\
a'&=\pm a f \sqrt{G-\beta ^2 w\over w},
\end{align}
which is again equal to the BPS equations \eqref{eq:GenBPSd} for dyon in the Generalized $SU(2)$ Yang-Mills-Higgs model.

\section{Discussion}
We have shown that the BPS Lagrangian method, which was used before in~\cite{Atmaja:2015umo} for BPS vortex, can also be applied to the case of BPS monopole and dyon in $SU(2)$ Yang-Mills-Higgs model (\ref{eq:canonicalLagrangian}). One main reason is because the effective Lagrangian (\ref{eq:effcanLag}) only depends on radial coordinate similar to the case of BPS vortex. We also took similar ansatz for the BPS Lagrangian (\ref{eq:LagBPS}) in which the BPS energy function $Q$ (\ref{eq:Q}) does not depend on the radial coordinate explicitly and it is a separable function of $f$ and $a$. This due to no explicit dependent over radial coodinate on the ansatz for the fields written in spherical coordinates as in (\ref{spherical ansatz}).

The BPS dyon could be obtained by identifiying the effective field of the time-component gauge fields $j$ to be propotional with the effective field of the scalars $f$ by a constant $\beta$, $j=\beta f$. This identification seems natural by realizing that both effective fields give the same Euler-Lagrange equation in the BPS limit. Fortunately we found that the BPS Lagrangian method forced us to take this limit when solving the last equation with explicit power of radial coodinate order by order, which are also the case for all other models considered in this article. In this article we used this simple identification which gives us the known result of BPS dyon~\cite{Prasad:1975kr}. It turns out that the constant $\beta$ takes values $|\beta|< 1$ and it will be equal to BPS dyon in~\cite{Prasad:1975kr,Coleman:1976uk} if we set $\beta=-\sin{\alpha}$, with $\alpha$ is a constant. There is also a possibility where the both effective fields are independent, or having no simple relation, but this will be discussed elsewhere.

Appliying the BPS Lagrangian method to Born-Infeld extensions of the $SU(2)$ Yang-Mills-Higgs model, which is called Nakumula-Shiraishi models, we obtained the same BPS equations as shown in~\cite{Nakamula:2013rfa}. Those BPS equations switch the sign if we shift the potential to a non-zero constant $4b^2$, $V\to V+ 4b^2$ in which the BPS limit now becomes $V\to 4b^2$, as shown below the equation (\ref{4b^2}). Therefore adding the energy density to a constant $4b^2$ seems to be related to a transition from monopole(dyon) to anti-monopole(anti-dyon) and vice versa. Since this transition is between BPS monopoles, or dyons, it would be interesting to study continuous transitions by adding the energy density slowly from $0$ to $4b^2$, which we would guest to be transition from BPS monopole(dyon) to Non-BPS monopole and then to the corresponding BPS anti-monopole(anti-dyon) with higher energy. This transition also appears in all Born-Infeld type of action discussed in this article and we wonder if this transition is generic in all other type of Born-Infeld actions at least with the ones posses BPS monopole(dyon) in the BPS limit. However, this kind of transition does not appear in $SU(2)$ Yang-Mills-Higgs model and its Generalized version since it would correspond to taking $b\to\infty$ in the Nakamula-Shiraishi models, which means adding an infinite potential energy to the Lagrangians.

In particular case of monopole Lagrangian (\ref{eq:NSmonopole}), we might try to use the identification $j=\beta f$, as previuosly, into the Lagrangian (\ref{eq:NSmonopole}) and look for the BPS equations for dyon from it. However, there is no justification for this identification because the Euler-Lagrange equations for $f$ and $j$ are not identical even after substituting $j=\beta f$ into both Euler-Lagrange equations in the BPS limit. We might also try to consider $f$ and $j$ independently by adding a term that is proportional to $j'$ in the BPS Lagrangian (\ref{eq:LagBPS}), but it will turn out that this term must be equal to zero and thus forces us to set $j=0$. Suprisingly, for the case of dyon Lagrangian (\ref{eq:NSdyon}), the effective action (\ref{eq:effNSd}) gives the identical Euler-Lagrange equations for $f$ and $j$ upon substituting $j=\beta f$ in the BPS limit. Therefore it is valid to use this identification for particular Born-Infeld type action of (\ref{eq:NSdyon}) for dyon.

We also applied the BPS Lagrangian method to the Generalized version of $SU(2)$ Yang-Mills-Higgs model (\ref{eq:LagcanG}) in which the effective action is given by (\ref{eq:gencanLag}). For monopole case, we found there is a constraint between the scalar-dependent couplings of gauge kinetic term $w$ and of scalar kinetic term $G$, which is $G=1/w$, similar to the one obtained in~\cite{Casana:2012un}. The BPS equations are also modified and depend explicitly on these scalar-dependent couplings. For dyon case, the constraint is generalized to $w(G-\beta^2w)=1-\beta^2$, with $\beta<|1|$, and the BPS equations are modified as well. This is relatively new result compared to~\cite{Casana:2012un,Casana:2013lna} in which they did not discussed about dyon. As previously assumed $w,G>0$, the constraint leads to $w_\pm={1\over 2\beta^2}\left(G\pm\sqrt{G^2-4\beta^2\left(1-\beta^2\right)}\right)$. Reality condition on $w_\pm$ gives lower bound to $G$ as such $G\geq|2\beta\sqrt{1-\beta^2}|$ in all values of radius $r$. The Generalized version of Nakamula-Shiraishi model for monopole, with Lagrangian (\ref{NSMonopole}) and effective Lagrangian (\ref{NSMonopoleEff}), has also been computed. The results are similar to the Generalized version of $SU(2)$ Yang-Mills-Higgs model for monopole in the BPS limit. In the case for Generalized version of Nakamula-Shiraishi model for dyon, with Lagrangian (\ref{NSDyon}) and effective Lagrangian (\ref{NSDyonEff}), the results are similar to the Generalized version of $SU(2)$ Yang-Mills-Higgs model for dyon, eventhough there are two additional scalar-dependent couplings $G_1$ and $G_2$. These additional couplings are related to the kinetic terms's couplings by $G_2-\beta^2 G_1=w(G-\beta^2 w)$. In the appendix, based on our results, we write down explicitly the complete square-forms of all effective Lagragians (\ref{eq:effNSm}), (\ref{eq:effNSd}), (\ref{NSMonopoleEff}), and (\ref{NSDyonEff}).


\acknowledgments

We would like to thank Handhika Satrio Ramadhan during the initial work of this article. A.N.A would like to thank CERN for hospitality during the visit, that was supported by RISET-PRO Non-degree 2017 program, where the initial writing of this article has been done.

\appendix
\section{Complete Square-Forms for Monopoles in Nakamula-Shiraishi Model}

For $V=0$, the effective Lagrangian (\ref{eq:effNSm}) can be rewritten in complete square-forms as the following:
\begin{eqnarray}
 \mathcal{L}_{\text{NSm}}&=&-\frac{b^2}{\sqrt{\left(1+\frac{1}{2 b^2}\left(\frac{2 a'^2}{r^2}+\frac{\left(a^2-1\right)^2}{r^4}\right)\right) \left(1+\frac{1}{2 b^2}\left(f'^2+\frac{2 a^2 f^2}{r^2}\right)\right)}}\times\nonumber \\
 &&\times\left(\frac{\left(2 b^2+f'^2\right)}{2 b^4 r^2}\left(a'- af\frac{\left(a^2-1\right) f'\pm 2 b^2 r^2}{r^2 \left(2 b^2+f'^2\right)}\right)^2+\frac{\left(2 a^2 f^2+r^2 \left(2 b^2+f'^2\right)\right)}{2 b^2 r^2 \left(2 b^2+f'^2\right)}\left(f'\mp\frac{a^2-1}{r^2}\right)^2\right.\nonumber\\
 & &\qquad +\left.\left(\sqrt{\left(1+\frac{1}{2 b^2}\left(\frac{2 a'^2}{r^2}+\frac{\left(a^2-1\right)^2}{r^4}\right)\right) \left(1+\frac{1}{2 b^2}\left(f'^2+\frac{2 a^2 f^2}{r^2}\right)\right)}-1\right.\right.\nonumber\\
 &&\qquad\qquad\left.\left.\mp\left(\frac{a^2-1}{2 b^2 r^2}f'+\frac{a f}{b^2 r^2}a'\right)\right)^2\right)\nonumber\\
&&\mp\left(\frac{2 a f}{r^2}a'+\frac{a^2-1  }{r^2}f'\right).
\end{eqnarray}
The above expression is different from the one presented in~\cite{Nakamula:2013rfa}.

For $V=4b^2$, it becomes
\begin{eqnarray}
 \mathcal{L}_{\text{NSm}}&=&-\frac{b^2}{\sqrt{\left(1+\frac{1}{2 b^2}\left(\frac{2 a'^2}{r^2}+\frac{\left(a^2-1\right)^2}{r^4}\right)\right) \left(1+\frac{1}{2 b^2}\left(f'^2+\frac{2 a^2 f^2}{r^2}\right)\right)}}\times\nonumber \\
 &&\times\left(\frac{\left(2 b^2+f'^2\right)}{2 b^4 r^2}\left(a'- af\frac{\left(a^2-1\right) f'\mp 2 b^2 r^2}{r^2 \left(2 b^2+f'^2\right)}\right)^2+\frac{\left(2 a^2 f^2+r^2 \left(2 b^2+f'^2\right)\right)}{2 b^2 r^2 \left(2 b^2+f'^2\right)}\left(f'\pm\frac{a^2-1}{r^2}\right)^2\right.\nonumber\\
 & &\qquad+\left.\left(\sqrt{\left(1+\frac{1}{2 b^2}\left(\frac{2 a'^2}{r^2}+\frac{\left(a^2-1\right)^2}{r^4}\right)\right) \left(1+\frac{1}{2 b^2}\left(f'^2+\frac{2 a^2 f^2}{r^2}\right)\right)}-1\right.\right.\nonumber\\
 &&\qquad\qquad\left.\left.\pm\left(\frac{a^2-1}{2 b^2 r^2}f'+\frac{a f}{b^2 r^2}a'\right)\right)^2\right)\nonumber\\
&&\pm\left(\frac{2 a f}{r^2}a'+\frac{a^2-1  }{r^2}f'\right).
\end{eqnarray}

Its general expression can be written as
\begin{eqnarray}
 \mathcal{L}_{\text{NSm}}&=&-\frac{2b^2\left(\left(\frac{V}{2 b^2}-1\right)^2+1\right)^{-1}}{\sqrt{\left(1+\frac{1}{2 b^2}\left(\frac{2 a'^2}{r^2}+\frac{\left(a^2-1\right)^2}{r^4}\right)\right) \left(1+\frac{1}{2 b^2}\left(f'^2+\frac{2 a^2 f^2}{r^2}\right)\right)}}\times\nonumber \\
 &&\times\left(\frac{\left(2 b^2+f'^2\right)}{2 b^4 r^2}\left(a'- af\frac{\left(a^2-1\right) f'\pm \left(2 b^2-V\right) r^2}{r^2 \left(2 b^2+f'^2\right)}\right)^2\right.\nonumber\\
 &&\qquad+\left.\frac{\left(2 a^2 f^2+r^2 \left(2 b^2+f'^2\right)\right)}{2 b^2 r^2 \left(2 b^2+f'^2\right)}\left(f'\mp\frac{a^2-1}{2b^2r^2}\left(2b^2-V\right)\right)^2\right.\nonumber\\
 & &\qquad+\left.\left(\left(\frac{V}{2 b^2}-1\right)\left(\sqrt{\left(1+\frac{1}{2 b^2}\left(\frac{2 a'^2}{r^2}+\frac{\left(a^2-1\right)^2}{r^4}\right)\right) \left(1+\frac{1}{2 b^2}\left(f'^2+\frac{2 a^2 f^2}{r^2}\right)\right)}-1\right)\right.\right.\nonumber\\
 &&\qquad\qquad\left.\left.\pm\left(\frac{a^2-1}{2 b^2 r^2}f'+\frac{a f}{b^2 r^2}a'\right)\right)^2\right.\nonumber\\
 &&\qquad+\left.\frac{V}{4 b^4}\left(4 b^2-V\right) \left(\frac{\left(a^2-1\right)^2}{2 b^2 r^4}+1\right) \left(\frac{2 a^2 f^2}{r^2 \left(2 b^2+f'^2\right)}+1\right)\right)\nonumber\\
&&\pm\frac{2 \left(\frac{V}{2 b^2}-1\right)}{\left(\frac{V}{2 b^2}-1\right)^2+1}\left(\frac{2 a f}{r^2}a'+\frac{a^2-1  }{r^2}f'\right)-\frac{V^2 \left(V-2 b^2\right)}{8 b^4-4 b^2 V+V^2},
\end{eqnarray}
which is valid only if $V=0$ or $V=4b^2$.

\section{Complete Square-Forms for Dyons in Nakamula-Shiraishi Model}
General expression for the complete square-forms of effective Lagrangian (\ref{eq:effNSd}) is given by
\begin{eqnarray}
\mathcal{L}_\text{NSd}&=&
-\frac{\left(2 b^2\right) \left(\left(1-\frac{V}{2 b^2}\right)^2+1\right)^{-1}}{\sqrt{1+\frac{\left(1-\beta ^2\right) \left(f'^2+\frac{2 a^2 f^2}{r^2}\right)}{2 b^2}+\frac{\left(\frac{\left(a^2-1\right)^2}{r^4}+\frac{2 a'^2}{r^2}\right)}{2 b^2}+\frac{\left(1-\beta ^2\right) \left(\frac{\left(a^2-1\right) f'}{r^2}+\frac{2 a a' f}{r^2}\right)^2}{4 b^4}}}\times\nonumber\\
&&\times\left(\frac{\left(1-\beta ^2\right)}{2 b^2}\left(f'\pm\frac{\left(a^2-1\right) r^2 \left(V-2 b^2\right)}{2 b^2 \sqrt{1-\beta ^2}  r^4}\right)^2
+\frac{1}{b^2 r^2}\left(a'\pm af\left(\frac{V}{2 b^2}-1\right)\sqrt{1-\beta ^2}
\right)^2
\right.\nonumber\\
&&\left.\qquad+\frac{V \left(4 b^2-V\right)}{8 b^6 r^4} \left(2 r^2 \left(a^2 \left(1-\beta ^2\right) f^2\right)+\left(a^2-1\right)^2+2 b^2 r^4\right)
\right.\nonumber\\
&&\left.\qquad+\left(\left(\frac{V}{2 b^2}-1\right)\sqrt{1+\frac{\left(f'^2+\frac{2 a^2 f^2}{r^2}\right)}{2 b^2\left(1-\beta ^2\right)^{-1}}+\frac{\left(\frac{\left(a^2-1\right)^2}{r^4}+\frac{2 a'^2}{r^2}\right)}{2 b^2}+\frac{\left(\frac{\left(a^2-1\right) f'}{r^2}+\frac{2 a a' f}{r^2}\right)^2}{4 b^4 \left(1-\beta ^2\right)^{-1}}}
\right.\right.\nonumber\\
&&\left.\left.\qquad\qquad +1-\frac{V}{2 b^2}\pm\sqrt{1-\beta ^2}\left( \frac{ af }{ b^2r^2}a'+\frac{ \left(a^2-1\right)}{2 b^2r^2}f'\right)\right)^2
\right)
\nonumber\\
&&\mp\frac{2 \left(1-\frac{V}{2 b^2}\right)\sqrt{1-\beta ^2}}{\left(1-\frac{V}{2 b^2}\right)^2+1}\left(\frac{ 2 af }{r^2}a'+\frac{\left(a^2-1\right)}{r^2}f'\right)
+\frac{V^2 \left(2 b^2-V\right)}{8 b^4-4 b^2 V+V^2},
\end{eqnarray}
which is valid only if $V=0$ or $V=4b^2$.

\section{Complete Square-Forms for Monopoles in Generalized Nakamula-Shiraishi Model}
General expression for the complete square-forms of effective Lagrangian (\ref{NSMonopoleEff}) is given by
\begin{eqnarray}
 \mathcal{L}_\text{NSmG}&=&
-\frac{2 b^2 \left(\left(1-\frac{V}{2 b^2}\right)^2+1\right)^{-1}}{\sqrt{\left(1+\frac{w }{2 b^2}\left(\frac{\left(a^2-1\right)^2}{r^4}+\frac{2 a'^2}{r^2}\right)\right) \left(1+\frac{G }{2 b^2}\left(\frac{2 a^2 f^2}{r^2}+f'^2\right)\right)}}\times\nonumber\\&&
\times\left(\frac{G \left(a'^2 w+b^2 r^2\right)}{2 b^4 r^2}\left(f'-\frac{2 a \left(a^2-1\right) a' f G w\pm\left(a^2-1\right) \sqrt{G w} r^2 \left(2 b^2-V\right)}{2 G r^2 \left(a'^2 w+b^2 r^2\right)}\right)^2
\right.\nonumber\\&&\left.
\qquad+\frac{w \left(w \left(\left(a^2-1\right)^2+2 a'^2 r^2\right)+2 b^2 r^4\right)}{2 b^2 r^4 \left(a'^2 w+b^2 r^2\right)}\left(a'\mp\frac{2 a f \sqrt{G w} \left(2 b^2-V\right)}{4 b^2 w}\right)^2
\right.\nonumber\\&&\left.
\qquad+V \left(4 b^2-V\right)\frac{\left(w \left(\left(a^2-1\right)^2+2 a'^2 r^2\right)+2 b^2 r^4\right) \left(a^2 f^2 G+b^2 r^2\right)}{8 b^6 r^4 \left(a'^2 w+b^2 r^2\right)}
\right.\nonumber\\&&\left.
\qquad+\left(\left(\frac{V}{2 b^2}-1\right)\left(\sqrt{\left(1+\frac{w \left(\frac{\left(a^2-1\right)^2}{r^4}+\frac{2 a'^2}{r^2}\right)}{2 b^2}\right) \left(1+\frac{G \left(\frac{2 a^2 f^2}{r^2}+f'^2\right)}{2 b^2}\right)}-1\right)
\right.\right.\nonumber\\&& \left.\left.
\qquad\qquad\pm\left(\frac{ a f \sqrt{G w}}{b^2r^2}a'+\frac{\left(a^2-1\right) \sqrt{G w}}{2b^2r^2}f'\right)\right)^2\right)
\nonumber\\&&
\mp\frac{2 \left(1-\frac{V}{2 b^2}\right) }{\left(1-\frac{V}{2 b^2}\right)^2+1}\left(\frac{2 a f \sqrt{G w}}{r^2}a'+\frac{\left(a^2-1\right) \sqrt{G w}}{r^2}f'\right)+\frac{V^2 \left(2 b^2-V\right)}{8 b^4-4 b^2 V+V^2},
\end{eqnarray}
which is valid only if $V=0$ or $V=4b^2$.

\section{Complete Square-Forms for Dyons in Generalized Nakamula-Shiraishi Model}
General expression for the complete square-forms of effective Lagrangian (\ref{NSDyonEff}) is given by
\begin{eqnarray}
\mathcal{L}_\text{NSdG}&=&
-\frac{2 b^2 \left(\left(1-\frac{V}{2 b^2}\right)^2+1\right)^{-1}}{\sqrt{1+\frac{\left(G-\beta ^2 w\right) \left(f'^2+\frac{2 a^2 f^2}{r^2}\right)}{2 b^2}+\frac{w \left(\frac{2 a'^2}{r^2}+\frac{\left(a^2-1\right)^2}{r^4}\right)}{2 b^2}+\frac{\left(G_2-\beta ^2 G_1\right) \left(\frac{\left(a^2-1\right) f'}{r^2}+\frac{2 a a' f}{r^2}\right)^2}{4 b^4}}}\times\nonumber\\&&
\times\left(\frac{(G-\beta ^2 w)}{2 b^2} \left(f'\mp\frac{\left(a^2-1\right) \sqrt{G_2-\beta ^2 G_1} \left(2 b^2-V\right)}{2 b^2 r^2 \left(G-\beta ^2 w\right)}\right)^2
\right.\nonumber\\&&\left.
\qquad+\frac{w}{b^2 r^2} \left(a'\mp\frac{\left(2 b^2-V\right) \left(4 a f\right)\sqrt{G_2-\beta ^2 G_1}}{8 b^2 w}\right)^2
\right.\nonumber\\&&\left.
\qquad+\left(\beta ^2 \left(G_1-w^2\right)-(G_2-G w)\right) \frac{\left(2 a^2 f^2 r^2 \left(G-\beta ^2 w\right)+\left(a^2-1\right)^2 w\right)}{8 b^6 r^4 w \left(G-\beta ^2 w\right)\left(V-2 b^2\right)^{-2} }
\right.\nonumber\\&&\left.
\qquad+V \left(4 b^2-V\right)\frac{ \left(2 a^2 f^2 r^2 \left(G-\beta ^2 w\right)+\left(a^2-1\right)^2 w\right)}{8 b^6 r^4}
\right.\nonumber\\&&\left.
\qquad+\left(\left(\frac{V}{2 b^2}-1\right)\sqrt{1+\frac{\left(f'^2+\frac{2 a^2 f^2}{r^2}\right)}{2 b^2 \left(G-\beta ^2 w\right)^{-1}}+\frac{\left(\frac{2 a'^2}{r^2}+\frac{\left(a^2-1\right)^2}{r^4}\right)}{2 b^2 w^{-1} }+\frac{\left(\frac{\left(a^2-1\right) f'}{r^2}+\frac{2 a a' f}{r^2}\right)^2}{4 b^4\left(G_2-\beta ^2 G_1\right)^{-1}}}
\right.\right.\nonumber\\&&\left.\left.
\qquad\qquad+1-{V\over 2b^2}\pm\left(\frac{a f \sqrt{G_2-\beta ^2 G_1}}{b^2r^2}a'+\frac{\left(a^2-1\right) \sqrt{G_2-\beta ^2 G_1}}{2b^2r^2}f'\right)\right)^2
\right)\nonumber\\&&
\mp\frac{2 \left(1-\frac{V}{2 b^2}\right)\sqrt{G_2-\beta ^2 G_1} }{\left(1-\frac{V}{2 b^2}\right)^2+1}\left(\frac{ 2 a f }{r^2}a'+\frac{\left(a^2-1\right)}{r^2}f'\right)+\frac{V^2 \left(2 b^2-V\right)}{8 b^4-4 b^2 V+V^2},
\end{eqnarray}
which is valid only if $V=0$ or $V=4b^2$, and $\beta ^2 \left(G_1-w^2\right)=G_2-G w$.

\end{document}